\documentclass[10pt]{article}

\usepackage{amsxtra,amssymb,amsthm,amsmath,latexsym}

\usepackage{graphicx}

\newcommand{\R}{{\mathbb R}}

\newcommand{\bee}{\begin{equation*}}
\newcommand{\eee}{\end{equation*}}
\newcommand{\be}{\begin{equation}}
\newcommand{\ee}{\end{equation}}
\newcommand{\pn}{\par\noindent}
\title{Creating desired potentials by embedding small inhomogeneities}
\author{A G Ramm\\
\small Department of Mathematics\\[-0.8ex]
\small Kansas State University, Manhattan, KS 66506-2602, USA\\[-0.8ex]
\small \texttt{ramm@math.ksu.edu}\\
}

\begin{document}
\date{}
\maketitle
\begin{abstract}
The governing
equation is $[\nabla^2+k^2-q(x)]u=0$ in $\R^3$. 
It is shown that any desired potential $q(x)$, vanishing outside a 
bounded domain $D$, can be obtained if one 
embeds into D many small scatterers $q_m(x)$, vanishing outside balls
$B_m:=\{x: |x-x_m|<a\}$, such that  $q_m=A_m$ in $B_m$,
$q_m=0$ outside  $B_m$, $1\leq m \leq M$, $M=M(a)$. 
It is proved that if the number of small scatterers in any subdomain $\Delta$
is defined as $N(\Delta):=\sum_{x_m\in \Delta}1$
and is given by the formula $N(\Delta)=|V(a)|^{-1}\int_{\Delta}n(x)dx [1+o(1)]$ as $a\to 0$,
where $V(a)=4\pi a^3/3$, then the limit of the function $u_{M}(x)$,
$\lim_{a\to 0}U_M=u_e(x)$ does exist
and solves the equation  $[\nabla^2+k^2-q(x)]u=0$  in $\R^3$, where $q(x)=n(x)A(x)$,
and $A(x_m)=A_m$.
The total number $M$ of small inhomogeneities is equal to $N(D)$ and is 
of the order $O(a^{-3})$ as $a\to 0$.

A similar result is derived in the one-dimensional case.

\end{abstract}

\pn{
}

\section{Introduction}
Consider the scattering problem:
\be\label{e1} [\nabla^2+k^2-q(x)]u=0\quad in\quad \R^3,\quad
k=const>0, \ee 
\be\label{e2} u=e^{ik\alpha\cdot
x}+A(\beta,\alpha,k)\frac{e^{ikr}}{r}+o\left(\frac{1}{r}\right),\quad
r:=|x|\to \infty,\quad \beta=\frac{x}{r},\quad \alpha\in S^2, \ee
where $S^2$ is the unit sphere in $\R^3$, and
$A(\beta,\alpha,k)=A_q(\beta,\alpha,k)$ is the scattering amplitude
corresponding to the potential $q(x),$ $\alpha$ is the direction of the
incident plane wave, $\beta$ is a direction of the scattered wave,
and $k^2$ is the energy.

 Let us assume that $p=p_M(x)$ is a
real-valued compactly supported bounded  function, which is a sum of small inhomogeneities:
$p=\sum_{m=1}^M q_m(x)$, where $q_m(x)$ vanishes outside the ball
$B_m:=\{x: |x-x_m|<a\}$ and  $q_m=A_m$ inside $B_m$,
$1\leq m \leq M$, $M=M(a)$.  

The problem, we are studying in this paper, is:

{\it Problem P: 
Under what conditions the field $u_M$, which solves the
Schroedinger equation with the potential $p_M(x)$, has
a limit $u_e(x)$ as $a\to 0$, and this limit  $u_e(x)$ solves the Schroedinger equation with a 
desired potential $q(x)$?
}

We give a complete answer to this question. Theorem 1 (see below) is
our basic result.

Our answer is, basically, as follows: 

{\it Given an arbitrary potential $q(x)$, vanishing outside of an arbitrary large but finite
domain $D$, one can find a function $A(x)$ and a function $n(x)\geq 0$, such that $A(x_m)=A_m$,
$A(x)n(x)=q(x)$, and the limit $u_e(x)$ of $u_M(x)$ as $a\to 0$ does exist, and solves problem
\eqref{e1}-\eqref{e2}.
}

The notation $u_e(x)$ stands for the effective field, which is the limiting field in the medium.

The field $u_M$ is the unique solution to the integral equation:
\be\label{e3} 
u_M(x)=u_0(x)-\sum_{m=1}^M \int_D g(x,y,k)q_m(y)u_M(y)dy, 
\qquad g(x,y,k)=\frac {e^{ik|x-y|}}{4\pi |x-y|},
\ee
where $u_0(x)$ is the incident field, which one may take as the plane wave, for example, 
$u_0=e^{ik\alpha \cdot 
x}$, where $\alpha \in S^2$ is the direction of the propagation of the incident wave.

We assume that the scatterers are small in the sense $ka<<1$. Parameter $k>0$ is assumed fixed,
so the limits below are designated as limits $a\to 0$, and condition $ka<<1$ is valid as $a\to 
0$.
 
If $ka<<1$, then the following transformation of \eqref{e3} is valid:
\be\label{e4}
u_M(x)=u_0(x)-\sum_{m=1}^M \frac {e^{ik|x-x_m|}}{4\pi}A_m u_M(x_m) 
\int_{|y-x_m|<a}\frac {dy}{|x-y|}[1+o(1)]. 
\ee
In \eqref{e4} we have used the folowing simple estimates:
$$|x-x_m|-a\leq |x-y|\leq |x-x_m|+a, \qquad |y-x_m|\leq a.$$
These estimates imply that 
$e^{ik |x-y|}=e^{ik |x-x_m|}[1+o(1)]$ if $|y-x_m|<a$ and $a\to 0$.

We want to prove that the sum in \eqref{e4} has a limit as $a\to 0$, and
to calculate this limit assuming that the distribution of small
inhomogeneities or, equivalently, the points $x_m$, is given by formula
\eqref{e5}, see below, and $M=N(D)$, where $N(\Delta)$ is defined in
\eqref{e5} for any subdomain $\Delta \subset D$, and $N(D)$ is $N(\Delta)$
for $D=\Delta$.

Our basic new tool is the following lemma.

{\bf Lemma 1.} {\it If the points $x_m$ are distributed in a bounded
domain $D\subset \R^3$ so that their number in any subdomain
$\Delta\subset D$ is given by the formula \be\label{e5}
N(\Delta)=|V(a)|^{-1}\int_{\Delta}n(x)dx [1+o(1)]\qquad a\to 0, \ee where
$V(a)=4\pi a^3/3$, and $n(x)\geq 0$ is an arbitrary given continuous in
$D$ function, and if $f(x)$ is an arbitrary given continuous in $D$
function, then the following limit exists: \be\label{e6}
 \lim_{a\to 0}\sum_{m=1}^M f(x_m)V(a)=\int_Df(x)n(x)dx.
\ee
}

Let us state our basic result.
 
{\bf Theorem 1.} {\it If the small inhomogeneities are distributed so that 
\eqref{e5} holds, and $q_m(x)=0$ if $x\not\in B_m$, $q_m(x)=A_m$ if $x\in B_m$
where $B_m=\{x: |x-x_m|<a$, $A_m:=A(x_m)$, and
$A(x)$ is a given continuous in $D$ function, then 
the limit 
\be\label{e7}
 \lim_{a\to 0}u_M(x)=u_e(x)
\ee
does exist and solves problem \eqref{e1}-\eqref{e2} with 
\be\label{e8}
q(x)=A(x)n(x).
\ee
}

There is a large literature on wave scattering by small inhomogeneities. A recent paper is 
\cite{fm}. Our approach is new. Some of the ideas of this approach were 
earlier applied by the author to scattering by small particles embedded in an inhomogeneous 
medium (\cite{R509}-\cite{R190}).

In Section 2 proofs are given and the one-dimensional version of the result is formulated and proved.

\section{Proofs}

{\it Proof of Lemma 1.} Let $\{\Delta_p\}_{p=1}^P$ be a partition of $D$ into a union of small 
cubes $\Delta_p$ with centers $y_p$, without common interior points, and 
\be\label{e9}
\lim_{a\to 0} \max_p diam \Delta_p=0
\ee
One has:
\be\label{e10}
\sum_{m=1}^M f(x_m)V(a)=\sum_{p=1}^P f(y_p)V(a)\sum_{x_m\in \Delta_p}1 [1+o(1)].
\ee
We use formula \eqref{e5} and the assumption \eqref{e9}  and get
\be\label{e11}
\sum_{x_m\in \Delta_p}1=V(a) n(y_p)| \Delta_p| [1+o(1)],
\ee
where $| \Delta_p|$ is the volume of the cube $ \Delta_p$.

It follows from \eqref{e10} and  \eqref{e11} that
\be\label{e12}
\sum_{m=1}^M f(x_m)V(a)=\sum_{p=1}^P f(y_p)n(y_p)| \Delta_p| [1+o(1)],
\ee
which is the Riemannian sum for the integral in the right-hand side of  \eqref{e6},
and the assumption  \eqref{e9} allows one to write
\be\label{e13} 
f(x_m)=f(y_p)[1+o(1)]\qquad \forall x_m\in  \Delta_p,
\ee
if $f$ is continuous.

The Riemannian sum in  \eqref{e12} converges to  the integral in the right-hand side of  
\eqref{e6} provided that the function $f(x)n(x)$ is continuous, or, more generally,
it is bounded and its set of discontinuity points is of Lebesgue measure zero.

Lemma 1 is proved. \hfill $\Box$

{\it Proof of Theorem 1.} We apply Lemma 1 to the sum in \eqref{e4},
in which we choose $A_m:=A(x_m)$, where $A(x)$ 
is an arbitrary continuous in $D$ function
which we may choose as we wish. A simple calculation 
yields the following formula:
\be\label{e14}
\int_{|y-x_m|<a}|x-y|^{-1}dy=V(a)|x-x_m|^{-1}, \qquad |x-x_m|\geq a,
\ee
and 
\be\label{e15}
\int_{|y-x_m|<a}|x-y|^{-1}dy=2\pi(a^2- \frac{|x-x_m|^2}{3}), \qquad 
|x-x_m|\leq a.
\ee
Therefore, the sum in  \eqref{e4} is of the form  \eqref{e6} with
$$f(x_m)=\frac {e^{ik|x-x_m|}}{4\pi|x-x_m|}A(x_m)u_M(x_m)[1+o(1)].$$ 
Applying Lemma 1,
one concludes that the limit $u_e(x)$ in  \eqref{e7} does exist and solves the
integral equation
\be\label{e16}
u_e(x)=u_0(x)-\int_D \frac {e^{ik|x-y|}}{4\pi|x-y|}q(y)u_e(y)dy,
\ee
where $q(x)$ is defined by formula \eqref{e8}.

Applying the operator $\nabla^2 +k^2$ to  \eqref{e16}, one verifies that
the function  $u_e(x)$ solves problem  \eqref{e1}- \eqref{e2}. 

Theorem 1 is proved. \hfill $\Box$

{\it Remark 1.}  Our method can be applied to the one-dimensional 
scattering
problem. The role of the balls $B_m$ is now played by the segments: 
$B_m:=\{x: x\in \R^1, |x-x_m|<a\}$, the role of $D$ is played by an interval
$(c,d)$, the $V(a)=2a$ in the  one-dimensional case, an analog of formula 
\eqref{e5} for the number of small inhomogeneities 
$N(\Delta)=\sum_{x_m\in \Delta}1$ is:
\be\label{e17}
N(\Delta)=(2a)^{-1}\int_{\Delta} n(x)dx [1+o(1)],
\ee
and $\Delta$ is now any interval on the line. The total number $M$ 
of small inhomogeneities is 
now of the order of $O(a^{-1})$.

In the  one-dimensional case an analog of the function $g(x,y,k)$
is
\be\label{e18} 
g(x,y,k)=-\frac {e^{ik|x-y|}}{2ik}.
\ee 
An analog of the potential $q_m$ is
$q_m(x)=A_m$ inside the interval  $B_m$, $q_m(x)=0$ outside $B_m$, and
we assume that $A_m=A(x_m)$, where $A(x)$ is a continuous function
which we can choose at will. With these notations one can use equation \eqref{e4} 
without any change, but remeber that $g(x,y,k)$ is now defined as in \eqref{e18}.
An analog of \eqref{e4} now is:
\be\label{e19}
u_M(x)=u_0(x)+\sum_{m=1}^M \frac {e^{ik|x-x_m|}}{2ik}A(x_m) u_M(x_m)2a 
[1+o(1)].
\ee

An analog of Theorem 1 can be stated as follows:

{\bf Theorem 2.} {\it  If the small inhomogeneities are distributed so that
\eqref{e5} holds, and $q_m(x)=0$ if $x\not\in B_m$, $q_m(x)=A_m$ if $x\in B_m$
where $B_m=\{x: |x-x_m|<a$, $A_m:=A(x_m)$, and
$A(x)$ is a given continuous in $D$ function, then
the limit $u_e(x)$ in \eqref{e7}
does exist and solves problem \eqref{e1}-\eqref{e2} with
$q(x)$ defined in \eqref{e8}, $\nabla^2 u$ replaced by $
u^{\prime \prime}$, and the radiation condition
\eqref{e2} modified to fit the one-dimensional problem.
}

\end{document}